\documentclass[aps,prc,twocolumn,floatfix,showpacs,a4paper,nofootinbib]{revtex4-1}

\usepackage{graphicx} 
\usepackage{dcolumn} 
\usepackage{bm}         

\newcommand{\be}{\begin{equation}}
\newcommand{\ee}{\end{equation}}
\newcommand{\ba}{\begin{eqnarray}}
\newcommand{\ea}{\end{eqnarray}}
\newcommand{\bd}{\begin{displaymath}}
\newcommand{\ed}{\end{displaymath}}

\begin{document}

\title{New method for measuring longitudinal fluctuations and directed flow
in ultrarelativistic heavy ion reactions}

\author{
L.P.~Csernai$^{1,2}$, G. Eyyubova$^3$ and V.K.~Magas$^4$} 

\affiliation{
$^1$ Institute of Physics and Technology, University of Bergen, Allegaten 55, 5007 Bergen, Norway \\
$^2$ Wigner Research Centre for Physics, 1525 Budapest, Hungary\\
$^3$ Department of Physics, University of Oslo, POB 1048 Blindern, 0316 Oslo\\
$^4$ Departament d'Estructura i Constituents de la Mat\`eria, Universitat de Barcelona, 08028 Barcelona, Spain
}

\date{\today}

\begin{abstract}
It has been shown in recent ALICE@LHC measurements that the odd flow 
harmonics, in particular a directed flow, $v_1$, occurred to be weak 
and dominated by random fluctuations.
In this work we propose a new method, which makes the measurements 
more sensitive to the flow patters showing global collective symmetries. 
We demonstrate how the 
longitudinal center of mass rapidity fluctuations can be identified, and 
then the collective flow analysis  
can be performed in the event-by-event center of mass frame. Such a method 
can be very effective in separating the flow patterns originating from 
random fluctuations, and the flow patterns originating from the
global symmetry of the initial state. 
\end{abstract}



\maketitle

Directed flow was the first and most dominant flow pattern in early 
relativistic heavy ion studies \cite{v1-evolution}. It provided a solid base 
for the subsequent
more detailed works with higher multipole components.  With increasing beam 
energy the magnitude of the directed flow decreased as the
longitudinal momentum increased considerably, and this made the measurement of 
this flow component more difficult. 

Recently the strong elliptic flow was demonstrated at LHC, exceeding all 
measurements at lower energies for peripheral collisions \cite{ALICE-Flow1}. 
At the same time recent measurements indicate weak directed flow, dominated
by fluctuations \cite{QM2011}.

Nevertheless the experience from early studies tells us that identification 
of this flow component is possible and it can be separated from a strongly 
fluctuating
background. Our goal in this article is to propose a new method,
which makes the analysis of the flow patterns to be more sensitive to global 
collective symmetries, by identifying and then correcting for the 
longitudinal center of mass (CM) rapidity fluctuations. The proposed 
idea is rather general and can be implemented by all the experimental teams, 
although the particular way of identifying the CM rapidity 
event-by-even will depend on the particular set-up.  

Collective flow is parametrized, by the azimuthal angle distribution 
around the beam axis via the expansion
\ba
\frac{d^3N}{dydp_td\phi} = \frac{1}{2\pi}\frac{d^2N}{dydp_t} \left[ 1 
+  2v_1(y,p_t) \cos(\phi) + \right.  \nonumber \\ 
\left. 2v_2(y,p_t) \cos(2\phi) + \cdot\cdot\cdot \  \right]
\ea
where $y$ is the rapidity and $p_t$ is the transverse momentum and
$\phi$ is the azimuth angle in the transverse plane with respect to 
impact parameter vector.  The functions $v_1(y,p_t), v_2(y,p_t), ...$
are the harmonic flow expansion parameters.

Global Collective flow patterns, which follow the global symmetries of the
reaction Event-by-Event (EbE) provide valuable information of the
overall pressure and transport properties of matter.

Random fluctuations, especially in the initial state, can also lead
to flow processes, these however, are not correlated with the global
collision geometry and the correlation of the major axis of the asymmetry
arising from the fluctuation may have no correlation with the reaction
plane at all \cite{ALICE-v3}.

The goal of the present work is to identify and separate the Global
Collective dynamical features from random fluctuations. We present
some of the steps of such reconnaissance for Global Collective flow in
high energy heavy ion reactions. This is particularly important for those
harmonic components which are weak and difficult to identify.

The odd harmonics are dominated by fluctuating flow components
which make it difficult to identify the weaker global collective
flow components. 
Just as the transverse plane fluctuations, the beam directed fluctuations
also modify the initial shape \cite{CMSS11,Cheng11}, the 
tilt of the longitudinal axis and 
similarly the CM position.  This issue is not discussed in the
literature, although it influences all odd harmonics in the $v_n$ expansion.

Recent LHC data indicate that the reaction plane can be measured with
forward and backward calorimeters where the projectile and 
target spectator residues are detected providing a reliable detection
of the global Event reaction Plane {EP} \cite{QM2011},
and the corresponding azimuthal angle, $\Psi_{EP}$.

To say something concrete, below we will mostly concentrate on the first 
flow component, $v_1$, although, as it was already mentioned, the method 
is rather general and interesting for all odd components. 
According to a recent estimate 
\cite{CMSS11}
the $v_1$ flow shows a strong mirror antisymmetric structure as a 
function of rapidity, but this is smoothed out 
by random fluctuations of the CM motion of the participants.
Furthermore
it was pointed out that fluctuations may cause turbulence i.e. random
rotation \cite{Urs}. Although, this was analyzed in the transverse
plane, similar fluctuating rotation may also appear in the reaction plane,
which may further soften the directed flow peak.

\section{Longitudinal Center of Mass Fluctuations}

The idea is straightforward, if the participant $y^{CM}$ or
$\eta^{CM}$ is strongly fluctuating, one can measure it experimentally,
and take the EbE CM into account when the odd flow components are
evaluated. More than two dozen years ago a similar idea was developed,
to estimate EbE the azimuthal angle of the reaction plane better, and
enable the directed flow analysis in low multiplicity and low acceptance
reactions also \cite{DO85}.

If the acceptance cowers a large fraction of the momentum space of
emitted particles, the initial CM of the system and the final measured
CM are nearly identical. The final measured CM might deviate from
the original one if a substantial part of particles are not detected,
especially if the not-detected particles are not evenly distributed
in the momentum space.

\subsection{Participant rapidity from emitted particles}

The total 4-momentum of all measured particles of one event is given by
\vskip -2mm
\begin{equation}
P = \sum_{\nu=1}^{M} p_\nu  \ ,
\label{Pdef}
\end{equation}
where $M$ is the measured multiplicity of the event.
$P$ can be measured accurately for the pseudo-rapidity
acceptance range of the detector, $|\eta| \le \eta_{max} $.
The arising Center of Mass (C.M.) rapidity is
\begin{equation}
y^{CM} = \frac{1}{2} \ln \frac{E+P_z}{E-P_z}.
\label{yCM}
\end{equation}

If we do not have a good mass resolution the determination of
$E_i$, and therefore of $E$, may become problematic, so only
the pseudo-rapidity of the CM can be determined:
\begin{equation}
\eta^{CM} = \frac{1}{2} \ln \frac{|\vec{P}| + P_z}{|\vec{P}|-P_z}.
\label{etaCM}
\end{equation}

If the rapidity
acceptance of the detector is limited, then the measured longitudinal single particle momenta are also constrained.  
On the other side, the transverse momenta are not constrained. 
For example at ALICE TPC detector the rapidity acceptance
of the detector is limited to $|y|<0.9$,  what means that the measured longitudinal nucleon momentum can not be more than $p_{||}=1$ GeV/c. At the same time the nucleon transverse momentum distribution peaks at $p_t=1.2$ GeV/c \cite{Schukraft-QM2011}, and a large part of the  
distribution extends to a few GeV/c. 
Thus this constraint of the acceptance would result in underestimating
the CM rapidity. 

The pseudorapidity distribution for peripheral
events was analyzed in a simple theoretical few source model
\cite{Astrid}
and for a detector with an acceptance of $|y|<0.9$  the
reduction in CM pseudorapidity fluctuation was estimated to be
$
|\eta^{CM}| < 0.2 \ .
$
This is an estimate relevant for the ALICE TPC detector. In other detectors 
the acceptance might be wider, however, due to geometric 
limitations such an underestimation of $ |\eta^{CM}| $ may sill occur.
Thus, we attempt to estimate $\eta^{CM}$ in other way.

\subsection{Participant rapidity from spectators}

Let us consider that we have three subsystems:
(A) projectile spectators, (B) participants, and (C) target spectators.
We can measure the energies of A and C, $E_A$ and $E_C$, in the respective 
Zero Degree Calorimeters (ZDC): $ZDC_a$ and $ZDC_c$.  Then the energy 
and momentum conservation gives
\begin{eqnarray}
E_B &=& A_B\ m_{B\perp}\ {\rm cosh}(y^B) = E_{tot} - E_A - E_C\ ,
\nonumber \\
M_B &=& A_B\ m_{B\perp}\ {\rm sinh}(y^B) = - (M_A + M_C) 
\label{EMcons}
\end{eqnarray}

For example, at the present LHC Pb+Pb
reaction with energy per Nucleon $\epsilon_{N} = 1.38$ TeV/nucleon, 
the beam rapidity is $y_0=7.986$ and 
\begin{eqnarray}
E_{tot} &=& 2 A_{Pb}\, m_N\ {\rm cosh}(y_0)\,, \nonumber
\end{eqnarray}
where $m_N = 938.8\ {\rm MeV/c}^2$.

Furthermore the equations
\begin{eqnarray}
E_A &=& A_P\, m_N\ {\rm cosh}(y_0),
\nonumber \\
E_C &=& A_T\, m_N\ {\rm cosh}(-y_0),
\nonumber 
\end{eqnarray}
give the spectator numbers, $A_P$ and $A_T$, and
\begin{eqnarray}
M_A &=& A_P\, m_N\ {\rm sinh}(y_0),
\nonumber \\
M_C &=& A_T\, m_N\ {\rm sinh}(-y_0), 
\nonumber
\end{eqnarray}
as well as the mass number of subsystem $B$:  
$$
A_B = 2 A_{Pb} - A_P - A_T \ .
$$ 
Thus for an event,
dividing the second of eq. (\ref{EMcons}) by the first we can determine the
rapidity of subsystem $B$, which should be close to
the rapidity of the participant system. 
\begin{equation} 
y^{CM}_E \approx y^B =  {\rm artanh}%
\left(\frac{M_A + M_C}{E_{tot} {-} E_A {-} E_C}\right) \ .
\label{ycmE}
\end{equation}

Our system $B$ includes high energy "pre-equilibrium" particles,
which are not detected by the ZDCs, and do not form a
locally equilibrated system. To separate these two components
from one-another would need more information, and
a quantitative definition.

The part of the neutrons of the colliding nuclei, 
which are originating from the spectators and thus flying with
beam rapidity (and energy) can be detected by the forward and backward
neutron
Zero Degree Calorimeters.
These ZDCs return the total energy of the measured neutrons.
The ZDCs are between the two (Projectile and Target) beam pipes.
The colliding nuclei propagate before and after passing through the
point of intersection in these beam pipes. Magnetic fields deflect these 
and other charged fragments, so that only the uncharged neutrons end 
up in the ZDCs.

In central collisions the spectators contain very few, mainly
single nucleons, and thus ZDC energies tend to zero
on both sides, A and C.

At higher impact parameters, the spectators will become more 
massive and the energy deposited in the ZDCs is increasing. At
peripheral collisions, two residue spectators are expected at
opposite sides of the participant zone. These may contain single
protons and neutrons as well as bound nuclear fragments. Of these
only the single neutrons reach the ZDCs, because all charged fragments
are deflected away from the the joint beam direction (where the two colliding 
beams interact, and then are separated into the two beam pipes).

In the LHC Pb+Pb reactions at 1.38 GeV/nucleon beam energy in each beam,
results in a total neutron energy including both of the colliding
Pb nuclei
$E^n_{tot} = 2 N_{Pb}\, m_N\ {\rm cosh}(y_0)\ = 348\ $TeV, 
where 
$2 N_{Pb} = 2\times 126$ 
is the total neutron number in the collision.
The neutron ZDCs measure only single neutrons. At zero impact parameter
we do not expect any spectators, while at the maximum impact parameter,
$ b = 2 R_{Pb} \,,  $
we do not expect any single neutrons, as the two spectators are 
two whole Lead nuclei. At very small impact parameters the number of 
single neutrons are proportional to the number of spectators
$N_n(b)= (N/A) N_{Spect}(b)$, where the nucleon number of the spectators
in a collision at impact parameter $b$, can be obtained from the 
intersection geometry (like in the Glauber model). In the rest of
the impact parameter range the spectators undergo nuclear 
multi-fragmentation, which is a well studied area both
theoretically and experimentally
\cite{multifra}.
These multi-fragmentation studies enable us to estimate the
number of single neutrons for a given impact parameter $b$.

We can assume that at an intermediate impact parameter where
the single neutron number of spectators is near to its maximum, 
we have less than half of the total neutron number of the spectators.
Thus, we can assume the maximum of total ZDC energy of about  
$E^n_{tot} = 160\ $TeV 
In other words with this estimate from of the 2 $\times$ 126 
available neutrons,
at most about 2 $\times$ 58 single neutrons can reach the 
ZDCs.  With increasing impact parameter the number of bound
nuclear fragments in the spectators increases, and thus these events
are populating again the lower $E_{ZDC}$ domains.

This feature can be checked by plotting  
$E^n_{tot} = E_{ZDCa} + E_{ZDCc}$ as a function of the 
total charged multiplicity, $N_{ch}$ measured,
in the TPC. As discussed above 
the measured $E^n_{tot}$ first increases with decreasing $N_{ch}$,
then reaches a maximum at 
$N_{ch}^{crit} \approx 100-200$ 
and then it decreases again.  
This enables us to separate the
discussion of medium-peripheral collisions with multiplicity
below $N_{ch}^{crit}$ and extreme-peripheral collisions with
multiplicity higher than $N_{ch}^{crit}$. 
And thus we can separate the correlations between 
$E_{ZDCa}$ and $E_{ZDCc}$
for different multiplicity or centrality bins.

For example this feature of the ZDCs for the ALICE ZDCs is discussed 
in ref. \cite{Monteno}
in an early simulation, where incomplete spectator fragmentation is 
taken into account, and HIJING is used as event generator. For the
estimates a model based on the ALADIN Au-Au data and deuterons are
estimated from NA49 data.  
The correlation between the energy in the Zero Degree Calorimeters, 
versus the number of single spectator neutrons, based
on theoretical model estimates raised monotonically up to about 100
neutrons and then it dropped rapidly. This rapid change was
clearly attributed to the fact that at higher impact parameters
the measured neutron number does not include neutrons in bound fragments, 
and thus the ZDC energy peaks at a critical neutron multiplicity 
\cite{Monteno}.

The charged particle multiplicity does not provide a good reference
measure for peripheral collisions, due to small multiplicities and
therefore large fluctuations. As it is introduced in ref. 
\cite{ALICE-Flow1} we can have better visualization of the 
spectator energy, if we plot the total ZDC energy,
$E^n_{tot} = E^n_A + E^n_C$, against the event centrality
percentage. 

The different presentations of these data are connected by the 
definition of centrality bins, which depend on the
detector acceptance and the treatment of the lowest
and highest multiplicities. The connection is presented in
Fig. 1 of ref. \cite{ALICE-Flow1}. 

Now the above considerations and evaluation of $y_{CM}$ should be
reconsidered due to the fact that the ZDCs measure single
neutrons only and the number of single neutrons depends in a special
way from the impact parameter due to the formation of nuclear
fragments in the spectators or due to the incomplete dissociation
of the bound spectator nuclear fragment in to single nucleons.

We can do a quantitative estimate of $y_{CM}$, if we
conclude on the total spectator energy carried by all neutral and
charged fragments together. Such a correction can be based on the
measured $N_{ch}$ and $N_{ch}^{crit}$.  

\begin{figure*}
 \centering
 \includegraphics[width=7in]{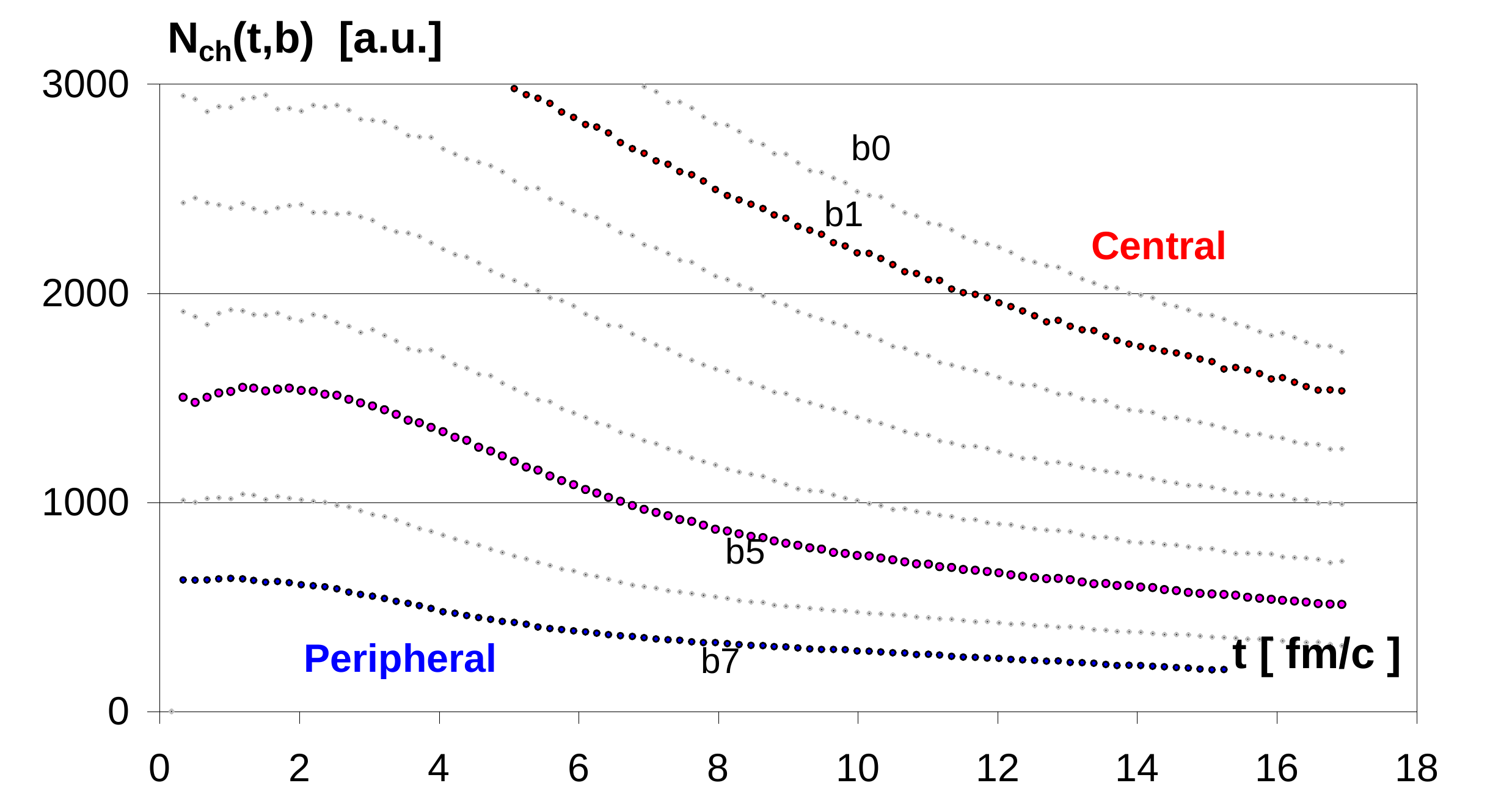}
 \caption{ 
Simulation of the  Pb+Pb collision at LHC with $\epsilon_{N} = 1.38$ TeV/nucleon \cite{CMSS11}. 
The calculated charged particle multiplicity,  $N_{ch}$, as a function of 
FO time (assuming a $t_{FO}=const.$ FO hyper-surface), for different 
impact parameters, $b=0.0, 0.1, 0.2, ... 0.7 b_{max}$. The indicated 
(b0, b1, ... b7) FO times for different impact parameters reproduce 
the measured charged particle multiplicities, $N_{ch}$, in the 
corresponding centrality bins. The visible fluctuations arise from
the feature of the PIC method \cite{CMSS11}, that the volume increases by one cell 
when a marker particle crosses the boundary. Thus at the initial state
with relatively few cells and large relative surface, this leads to
fluctuations.
}
 \label{f3}
\end{figure*}

We may attribute the charged particle multiplicities or the 
centrality percentage to the obtained multiplicity estimated
in a fluid dynamical model calculation, see Figure \ref{f3}, which is based on the calculations discussed in details in \cite{CMSS11}. 
The impact parameters of the fluid dynamical model calculation 
can be matched well to the centrality percentages.
The correspondence between impact parameter and event centrality percentage 
is impressive for semi-peripheral reactions: $b= 
0, 2, 0.3, 0.4, 0.5, 0.6, 0.7 b_{max} $ correspond to
   5,  13,   21, 29,  37,  45 \% centrality percentage respectively.

We should correct the preliminary estimate, eqs. (\ref{EMcons},\ref{ycmE}),
for the CM rapidity. We have to correct for the fact that nuclear 
fragments and the energy carried by these are not measured by the $ZDCs$,
and thus our estimates for $E_A, E_C$ and thus $E_B$ should be modified
and should be calculated based on the neutron energies measured in the 
ZDCs.

\subsubsection{Neutrons in nuclear fragments}

The {\bf simplest approximation} is that the total spectator energies,
$E_A$ and $E_C$ are related the same way to the measured ZDC energies:
$$
E_A = (A/N) E^n_A \ , \ \ \ \
E_B = (A/N) E^n_B \ , 
$$
where $A/N$ is the mass number to neutron number ratio in colliding 
nuclei (for simplicity we consider symmetric collisions).
This approximation is satisfactory for central collisions and for
small impact parameters (i.e. $b<0.3b_{max}$ so, relatively large 
charged multiplicities). 
Due to the presence of bound nuclear fragments in the spectators, we need a 
{\bf better approximation}
at intermediate or higher impact parameters.

\begin{figure}[h]
 \centering
 \includegraphics[width=3.5in]{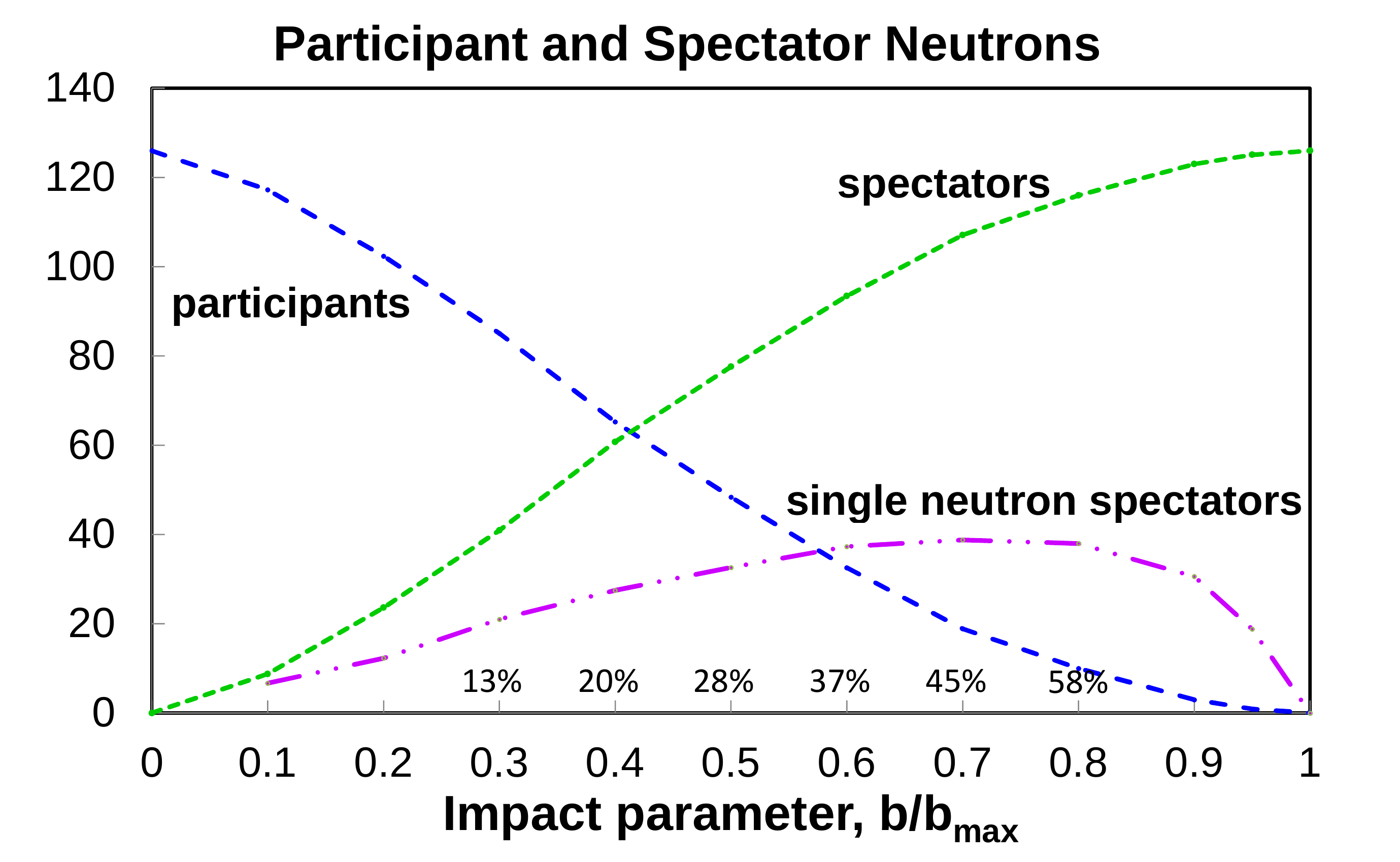}
 \caption{   
(color online)
The number of participant neutrons from the projectile or from the
target (dashed blue line) and the corresponding number of spectator 
neutrons (dotted green line) in one of the
spectators (forward or backward) obtained in the initial state
calculation \cite{MCs001} for Pb+Pb collision.. 
The neutron distribution is assumed to
be homogeneous in the system, $N/A=126/208$. At large impact parameters
part of spectator neutrons remain in nuclear fragments, which are
charged and so do not reach the neutron ZDCs. Based on the FD estimates
we relate impact parameter with the centrality percentage, and based on nuclear multi-fragmentation studies 
\cite{multifra},  see Table \ref{T1},
we estimate the number of single nucleons, which reach the
neutron ZDCs (magenta dashed double dotted line).
}
\label{n-vs-b}
\end{figure}

Based on the initial state model \cite{MCs001}, used in the 
3 dimensional fluid dynamical calculations \cite{CMSS11,CC10}
from the geometrical overlap of the colliding nuclei, we get
the number of participant nucleons  for each impact parameter.
On the projectile (or target) side 126/208 part of these are 
participant neutrons, $P_n(b)$, 
shown by dashed blue line in Figure \ref{n-vs-b}. 
The remaining neutrons are in spectators, shown by the dotted green 
line.  As in ref.
\cite{MCs001}, we assume that the N/Z ratio is homogeneous in the
whole initial state system.

Not all spectator neutrons are single neutrons; some of these
are in composite nuclear fragments. The total  energy of single neutron
spectators are measured in the ZDCs, and thus their number can be
easily obtained as the beam energy per nucleon is known.
So, we can get the number of single neutrons, $N_n(b)$,
as a function of the impact parameter $b$ in a straightforward way
from the total {\bf measured} ZDC neutron energy 
$E_{ZDCa} + E_{ZDCc} = E^n_{tot}$ as a function of the impact parameter
or the centrality percentage. The asymptotic behaviour of
this  dependence is not straightforward, $ E^n_{tot} \longrightarrow 0$
both if $b/b_{max} \longrightarrow 0$ or 1. 
Based on nuclear multi-fragmentation studies mentioned above 
\cite{multifra},
we can estimate the number of single spectator nucleons, $ N_n(b) $,
as given in
Table \ref{T1}. This estimate indicates that the number of 
detected single neutrons should have a maximum of about 30\% 
at $\sim $ 50\% centrality percentage.

\begin{table}
\begin{tabular}{ccccc}  \hline
$b/b_{max}$&centrality&$\langle N_n(b) \rangle$&$P_n(b)$ \\ 
      &        [\%]  &    &           \\ \hline
      & 0-10   &     6.7  &     8.8       \\
(0.2) &  6     &    12.3  &    23.5       \\ 
0.3   & 13$\pm$5 &  21.0  &    40.9       \\ 
0.4   & 20$\pm$5 &  27.5  &    60.8       \\ 
0.5   & 28$\pm$5 &  32.6  &    77.6       \\ 
0.6   & 37$\pm$5 &  37.3  &    93.4       \\ 
0.7   & 45$\pm$5 &  38.8  &   107.1       \\
(0.8) & 58$\pm$5 &  38.0  &   116.0       \\ 
(0.9) & 72$\pm$5 &  30.6  &   123.0       \\ 
(0.95)&(84$\pm$5)&  18.8  &   124.6       \\ 
(1.0) &(90$\pm$5)&   0.0  &   126.0       \\   \hline
\end{tabular}
\caption{The number of single spectator neutrons as function of 
centrality bins and the corresponding impact parameters as estimated based on 
nuclear multi-fragmentation model \cite{multifra}. The initial geometrical spectator numbers
corresponding to a given impact parameter are also given.}
\label{T1}
\end{table}

The difference between the average and maximum for the single
neutron numbers leads to an estimate of the systematic error of the
method, for estimating the total spectator energy and momentum from
the single neutrons.

Part of these spectator neutrons are inside composite nuclear fragments,
especially at large impact parameters. These are charged, and due
to the magnetic fields directing the beam, these do not reach the 
central neutron ZDCs nor the proton ZDCs. 
As discussed above his reduces the energy
detected in the ZDCs to about 130-140 TeV, and the maximum numbers of
detected single neutrons to about 47-51. This maximum appears at
the centrality percentage of $\sim$ 50\%, which corresponds to 
$b = 0.75 b_{max}$ according to our fluid dynamical model estimates.

Thus we estimate number of single neutrons
in the spectators for each impact parameter. These estimates have 
some systematic error due the unavoidable theoretical input,
and at very central collisions as well as for extreme peripheral
collisions (where there are no data and the multiplicity is so
low that the fluid dynamical approximation is not applicable). The
multifragmentation process may have a fluctuation itself, which may
interfere with the $E_A$ $E_C$ asymmetry, and lead to some systematic
error. 

In a given experimental setup one can {\it measure directly} $ N_n(b) $,
via measuring the average of the total energy in the two ZDCs as
function of the impact parameter, $b$, and dividing it by the 
beam energy per nucleon. This direct measurement of $ N_n(b) $,
should be preferred, as it eliminates large part of the uncertainties 
of the theoretical approach
arising from directly emitted high energy neutrons, and special 
emission evaporation mechanisms at the spectator/participant
boundary with large shear. 

Now we have to estimate, at a given centrality percentage
(or impact  parameter, $b$),
the total energy of the spectator residues (including protons and
charged fragments), $E_A$ and $E_C$, 
 from the energy of measured single neutrons in the
ZDCs, $E^{sn}_A$ and $E^{sn}_C$. From the measured spectator
neutron energies, $E^{sn}_A$ and $E^{sn}_C$, we
get the corresponding single neutron numbers, 
$N^{sn}_A = E^{sn}_A/\epsilon_{N}$ and 
$N^{sn}_C = E^{sn}_C/\epsilon_{N}$. Multiplying these numbers by 
the correction factor, $P_n(b)/N_n(b)$, we get the spectator numbers,
$A_{P,T} = N^{sn}_{A,C} P_n(b)/N_n(b)$, and spectator
energies, including the contributions of single protons and of
all nucleons bound in composite nuclear fragments:
\begin{eqnarray}
E_A(b) &=& (A/N) E^{sn}_A P_n(b)/N_n(b) \nonumber \\
E_C(b) &=& (A/N) E^{sn}_C P_n(b)/N_n(b) \,.
\end{eqnarray}
This yields the corresponding spectator momenta, $M_A, M_C$,
and we can get the event by event C.M. rapidity as in eq. 
(\ref{ycmE})
\begin{equation} 
y^{CM}_E(b) \approx y^B =  {\rm artanh}%
\left(\frac{M_A + M_C}{E_{tot} {-} E_A {-} E_C}\right) 
- y^{CM}(b)\ ,
\label{ycmZ}
\end{equation}
where the last term is added to correct for eventual
detector asymmetry, which is measurable,
for all events of the sample for a given multiplicity percentage
bin.
Now, $E_A, E_C, A_P, A_T$ are estimated and in the estimate
we used the average $N_n(b)$, based on the estimated or eventually  
measured ZDC energies.

The correction increases for increasing impact parameter,
which leads to increased estimates for $y_{CM}$ fluctuations, on the
other hand the correction is also increases and this leads to
the possibilities of larger systematic errors. 

At extreme peripheral reactions spectators
appear in the form of bound, and so charged, nuclear fragments, which
are diverted by the magnetic field and are not detected. These provide 
more room for large C.M. rapidity fluctuation, but unfortunately
neither the ALICE TPC nor the ZDCs can measure these reactions
with the needed acceptance.

The above suggested analysis, by using the ZDC and charged multiplicity
information together may still provide the best estimate
for the longitudinal fluctuations.

\subsection{Participant rapidity fluctuations}


In central
collisions the ellipticity from random fluctuations can easily 
supersede the one from the global symmetry. This is indicated by
the recent flow measurements, where the axes of different higher
harmonics are uncorrelated, indicating that the source is random 
fluctuations, which are not correlated with the global reaction plane.

The global asymmetry may become dominant at higher impact parameters,
i.e. $b > 0.6 - 0.8 b_{max}.$ For example, using the observed multiplicity in the ALICE TPC the centrality
bins were defined as shown in Ref. \cite{ALICE-Flow1}. These centrality
bins can be assigned to impact parameter intervals, based
on theoretical and geometrical model assumptions.

For example, in Ref. \cite{CMSS11} simulations of the Pb+Pb collisions at LHC have been performed. 
The comparison of the multiplicity estimates of this model provide a correspondence
between the impact parameter and the centrality percentage, see Fig. \ref{f3}.

For the azimuthal flow asymmetries arising from random fluctuations,
we cannot expect a correlation between the reaction plane (as measured
by the spectator residues), and the asymmetry observed in the TPC. 
This also applies to the center of mass of the observed
particles, because the longitudinal fluctuations are also not correlated
with the spectator residues. Thus, we can expect a correlation mainly
at peripheral collisions, where the ellipticity arising from global 
symmetry may be significant and may exceed the one caused by random
fluctuations. Based on the above discussed correspondence between the
multiplicity and the impact parameter, we can expect to see global effects
at 50\% centrality bin or higher, i.e. in rather peripheral 
collisions. 

The collective global flow components appear at moderate momenta,
thus the CM determination should be sufficient for this component of the
flow. In the previous subsection two methods were presented to estimate
the C.M. rapidity. Using the TPC with the restricted in pseudorapidity acceptance limits 
the observed longitudinal momenta, while the transverse momenta are
not constrained. Due to this the TPC is expected to underestimate
the C.M. rapidity fluctuations.

The other method based on the single neutrons in the ZDCs is not 
constrained the same way. On the other hand one has to estimate
the total spectator energy and momentum from the single neutrons, and
this step requires some theoretical estimate which introduces a
systematic error. This is illustrated by the fact that the correction
factor,  $P_n(b)/N_n(b)$ can be based on the average 
single neutron multiplicity. $N_n(b)$ can also be obtained from the
theoretical estimates or the
measured data, where the total single 
neutron energy, measured in the two ZDCs, is divided by the beam energy 
per single neutrons, $\epsilon_{N}$.

The EbE data would show large fluctuations within each event among the 
emitted particles, exceeding the C.M. fluctuations considerably. We expect that compared
to these fluctuations the C.M. fluctuations are small.

\subsection{Correlation between the TPC and ZDC C.M. rapidities}

Before we start to study the correlations we have to remove the
average C.M. rapidity shift from the data as this arises from
asymmetries in detector acceptance. This is probably negligible
for the TPC but expected to be significant for the ZDC data. 

The azimuthal correlations of global collective flow 
are in correlation with the reaction plane, also the
longitudinal C.M. rapidity should be globally defined for the
flow pattern. In such a situation one would expect that
the TPC and ZDC rapidities should strongly correlate.

On the other hand if the longitudinal and azimuthal correlations 
arise from random fluctuations, which are not correlated with the
global symmetry axes then such a correlation for C.M. rapidities
measured at different rapidity ranges should not correlate.

We anticipate that at central and semi-central
collisions there is no significant correlation or anti-correlation
among the C.M. rapidities shown by the particles detected in the 
TPC or the ZDCs. In these collisions the correlations are arising 
dominantly from random fluctuations.

For peripheral collisions of 50 \% centrality percentage or more 
a weak positive
correlation is anticipated. 
Here the global collective asymmetry
(eccentricity) is so strong, that the global correlations start
to become competitive.

\section{Adjustment of the Center of Mass} 

Substantial initial state rapidity fluctuations will average
out all flow structures around the CM when the measurements
are assuming that the CM is identical with the pre-collision
CM of the given experiment (i.e. fixed to the Laboratory frame
in a colliding beam experiment with a symmetric, A+A, collision). 
   Odd components of global collective flow patterns, 
($v_1$, $v_3$, ...)
are Mirror Asymmetric (MA) around the real participant CM, so these
are severely effected by the EbE CM fluctuations \cite{CMSS11,Cheng11}.

Based on the above results, we suggest to use the CM rapidity, EbE,
determined from the ZDC data.
Let us assume that the CM rapidity is measured for 
each Event, ($E$):
$$
y^{CM}_E(b) = \frac{1}{2} \ln \frac{E + P_z}{E - P_z}.
$$
where $P = \sum_{\nu=1}^{M} p_\nu $ is the total measured 4-momentum
of event $E$.
Here we consider that all corrections are taken into account as
described in eq. (\ref{ycmZ}). 
Let us then shift each event  to its own CM 
by the measured $y^{CM}_E$ so that
each particle rapidity $y_i$ will be moved to 
\begin{equation}
y'_i = y_i -  y^{CM}_E(b) \ 
\label{yshift1}
\end{equation}
This transformation will not effect the azimuth angle of the
emitted particles, $\vec{p}_{\perp, i}$, nor $m_{\perp, i}$,
however, the longitudinal momentum and the energy will change to,
\begin{eqnarray}
p'_{z, i} &=& m_{\perp, i} \sinh ( y_i -  y^{CM}_E ) 
\nonumber \\
E'_{i}    &=& m_{\perp, i} \cosh ( y_i -  y^{CM}_E ) .
\nonumber
\end{eqnarray}
Consequently the EbE $v_n(p_\perp)$ will not change after adjusting
the center of mass rapidity (if we integrate over the full rapidity
range). Instead in a given event, $E$, this adjustment will contribute 
to a re-distribution  $v_n(y_i)$ to of  at another set of rapidities, $y'_i$.
Consequently the detector acceptance boundaries will also change EbE, 
and so one will have a continuous rapidity coverage even if the 
detectors have some rapidity acceptance gaps. Then, the flow harmonics
can be determined EbE averaging over {\bf all measured particles in the Event} 
\begin{equation}
v_n (y',p_\perp)_E = \langle \ \cos[n(\phi_i - \Psi_{EP})] \ \rangle_E \ ,
\end{equation}
and then one can make an average over {\bf all events in a centrality bin:}
\begin{equation}
v_n (y',p_\perp) = \langle \ v_n (y',p_\perp)_E \ \rangle \ .
\end{equation}
If the centrality bin is wide this two step averaging is essential
\cite{DMZ10}.

As usual,
one can also correct for fluctuations of the observed Event Plane azimuth
angle, $\Psi_{EP}$, compared to the real (pre-collision, but not measurable)
Reaction Plane, $\Psi_R$, as  $\Psi_R \neq \Psi_{EP}$, and so the observed
$v_n^{obs}$ must be corrected by dividing by the resolution of the event
plane \cite{PV98}.

So, using eqs. (\ref{ycmZ},\ref{yshift1}) we can perform this rapidity shift event-by-event. 
To eliminate a relatively
large rapidity binning 
of the sample the events could be distributed uniformly and randomly inside
each rapidity bin. Then the pseudo-rapidity distribution of $v_1$,
as measured in the EbE CM frame is expected to show a clear
$\pm y$ asymmetry as the longitudinal, $y^{CM}$ fluctuations are removed from
the data. The distribution this way may show a distinctly
antisymmetric distribution in rapidity, as required by the 
global symmetry of the event. 

In case of ALICE, for example, the TPC data are available
in terms of pseudorapidity and integrated over the $p_\perp$ acceptance 
of the TPC. In such a situation instead of eq. (\ref{yshift1}), 
we can use the approximation that the pseudorapidity, EbE,
is shifted by the CM rapidity:
\begin{equation}
\eta'_i = \eta_i -  y^{CM}_E(b) \ 
\label{etashift1}
\end{equation}

So, this way the collective part of the
flow may be identified.

We can quantify the identification of the collective symmetric
versus the random fluctuating contribution to the flow the following
way. We can evaluate the {\it odd} and {\it even}
components of the flow \cite{QM2011}:
\begin{eqnarray}
v_n^{odd}(y',p_\perp)\ &=& [ v_n (y',p_\perp) + v_n (-y',p_\perp) ]
\nonumber \\
v_n^{even}(y',p_\perp) &=& [ v_n (y',p_\perp) - v_n (-y',p_\perp) ]\,.
\nonumber
\label{od-ev}
\end{eqnarray}
We can use here also the approximation eq. (\ref{etashift1}) to evaluate the
symmetries of the data.

Initial states with global collective symmetry are fully Mirror Asymmetric (MA) states,
while fluctuating states may have both MA and Mirror Symmetric (MS) components. 
Then the two,  {\it odd} and {\it even} components
for $v_1(\eta')$ should show a distinct difference.
The original $v_1(\eta')$ component is expected to be very 
close to the {\it odd} component,
while the {\it even} component should be much
smaller in the central rapidity range. This would indicate that
 $v_1(\eta')$ is MA to a large extent, and the MS component is consistent with zero.

At zero impact parameter global symmetry does not result in any
global azimuthal
asymmetry, so all harmonic coefficients must vanish from global
collective origin and only random initial state fluctuations as
well as dynamical random fluctuations developing during the collision
may lead to azimuthal fluctuations. The recent ALICE analysis of
azimuthal flow asymmetries central collisions \cite{QM2011}, 
could identify flow harmonics from   $v_1$ to $v_8$ and shows a maximum
for $v_3$!

\section{Conclusions}

We presented a method how to 
analyze the directed flow data 
by considering the EbE longitudinal rapidity
fluctuation of the participants. The participant CM 
rapidity can be estimated both from the TPC  and the ZDC data.
The TPC data are constrained to the pseudorapidity acceptance range,
which impairs the CM rapidity estimate for peripheral reactions.
To obtain an estimate for peripheral reactions from the ZDC data we 
estimate the forward and backward spectator energies from the measured
single neutron energies. 
For peripheral reactions, where the azimuthal asymmetry of
overall global collective origin is expected to exceed those, which 
originate from random initial state fluctuations, we expect to find 
significant correlation between the TPC and ZDC estimates of the
CM rapidity of the participants, in the central rapidity range.

Using the ZDC data, which are not constrained by the limited acceptance,
we describe each event in its own CM frame,  and propose to evaluate the
directed flow from these shifted data.

The method of shifting the system origin event by event to the
rapidity of the participant CM rapidity is
effective in separating the flow patterns originating from
random fluctuations, and the flow patterns originating from the
global symmetry/asymmetry of the initial state.

\begin{acknowledgments}

This work is supported in part by the European Community-Research
Infrastructure Integrating Activity ``Study of Strongly
Interacting Matter'' (HadronPhysics2, Grant Agreement n. 227431)
under the Seventh Framework Programme of EU, and the work
of V.K.M. is partly supported by the contracts FIS2008-01661 from
MICINN (Spain), by the Ge\-ne\-ra\-li\-tat de Catalunya contract
2009SGR-1289. Gyulnara Eyyubova thanks for the hospitality of the
University of Bergen, where part of this work was done and for the
financial support from the Norwegian State Educational Loan Fund.

\end{acknowledgments}

%

\end{document}